\documentclass[aps,prl,twocolumn,amsmath,amssymb]{revtex4-1}
\usepackage{graphicx}
\usepackage{subfigure}
\usepackage{epsfig}
\usepackage{dcolumn}
\usepackage{bm}
\usepackage{color}
\usepackage{amsbsy}
\usepackage{hyperref}

\def\be{\begin{equation}}       \def\ee{\end{equation}}
\def\bea{\begin{eqnarray}}      \def\eea{\end{eqnarray}}
\def\ba{\begin{array}}
\def\ea{\end{array}}
\def\bnum{\begin{enumerate} }
\def\enum{\end{enumerate}}

\def\nn{\nonumber}

\def\=>{\Rightarrow}
\def\>{\rightarrow}

\def\eye2{Fathbb{I}}

\newcommand{\s}{{\sigma}}

\renewcommand{\>}{\rangle}

\begin{document}

\title{Hourglass semimetals with nonsymmorphic symmetries in three dimensions}
\author{Luyang Wang$^{1,2}$, Shao-Kai Jian$^{1}$, Hong Yao$^{1,3,4}$}
\email{yaohong@tsinghua.edu.cn}
\affiliation{$^1$Institute for Advanced Study, Tsinghua University, Beijing 100084, China\\
$^2$State Key Laboratory of Optoelectronic Materials and Technologies, School of Physics, Sun Yat-Sen University, Guangzhou 510275, China\\
$^3$State Key Laboratory of Low Dimensional Quantum Physics, Tsinghua
University, Beijing 100084, China\\
$^4$Collaborative Innovation Center of Quantum Matter, Beijing 100084, China
}

\begin{abstract}
It was recently shown that nonsymmorphic space group symmetries can protect novel \textit{surface} states with hourglass-like dispersions. In this paper, we show that such dispersions can also appear in the \textit{bulk} of three-dimensional (3D) systems which respect nonsymmorphic symmetries. Specifically, we construct 3D lattice models featuring hourglass-like dispersions in the bulk, which are protected by nonsymmorphic and time-reversal symmetries. We call such systems hourglass semimetals, as they have point or line nodes associated with hourglass-like dispersions. Hourglass nodal lines appear in glide-invariant planes, while hourglass Weyl points can occur on screw-invariant axes. The Weyl points and surface Fermi arcs in hourglass Weyl semimetals are stable against weak perturbations breaking those nonsymmorphic symmetries. Our results may shed light on searching for exotic Weyl semimetals in nonsymmorphic materials.
\end{abstract}
	
\date{\today}
\maketitle

{\it Introduction}.---Topological insulators and topological superconductors are novel quantum states of matter with gapless surface/edge states protected by time-reversal and/or particle-hole symmetries, which have attracted vast attentions in the past decade\cite{hasan-2010,qi-2011,moore-2010}. It was shown later that other symmetries such as translational and point-group symmetries may also protect exotic gapless surface/edge states in fermionic systems; for instance, topological crystalline insulators protected by point group symmetries were proposed\cite{teo-kane-2008,fu-2011,hsieh-2012, slager-2012,yao-2013,chiu-2013,ando-2014,shiozaki-2014,chiu-2016}. These are among the prototype examples of symmetry protected topological (SPT) phases\cite{gu-wen-2009,pollmann-2012,chen-2012,Ashvin-2013,senthil-2015}. Besides point-group symmetries such as discrete rotation, reflection, and inversion symmetry, space-group symmetries also include nonsymmorphic symmetries, i.e., glide reflections and screw rotations. It has been shown that nonsymmorphic space group symmetries can protect novel phases, including both insulating phases\cite{liu-2014,shiozaki-2015,shiozaki-2016,fang-2015,dong-2016,lu-2016,parameswaran-2013,watanabe-2015,po-2016,watanabe-2016, parameswaran-2015, pixley-2016} and semimetallic phases\cite{young-2015,wieder-2016PRB,kim-2016,chen-2016,zhao-2016,yang-2016}. Since 157 out of the 230 space groups are nonsymmorphic, it is of importance to study various topological phases of electrons protected by nonsymmorphic symmetries.

Recently, it was shown that the combination of glide reflection symmetry and time-reversal symmetry can protect hourglass-like dispersion on the surface of nonsymmorphic insulators, e.g. KHg$X$($X$=As, Sb, Bi), with a gapped bulk of nontrivial topology\cite{wang-2016,alexandradinata-2016,ma-2016,ezawa-2016,chang-2016}. The hourglass-like dispersion is intimately related to the M\"obius twist of the surface states\cite{shiozaki-2016,shiozaki-2015}. In contrast to the Dirac fermions that appear on the surface of topological insulators, the hourglass fermions have dispersions of four bands, with double degeneracy at high symmetry points of the Brillouin zone and an unavoidable band crossing between those high-symmetry points. As ordinary Dirac/Weyl fermions can exist both on the surface and in the bulk of three-dimensional (3D) systems, one can ask if Weyl fermions with hourglass-like dispersions can appear in the bulk of 3D crystals, besides on the surface.

In this work, we construct 3D lattice models that host fermions with hourglass-like bulk dispersions. Such dispersion is protected by time-reversal and glide reflection (or screw rotation) symmetries. When the Fermi energy is tuned to the neck of the hourglass, semimetals with point or line degeneracies result, which we name 3D hourglass semimetals. There are two distinct types of 3D hourglass semimetals: hourglass Weyl semimetals (HWSMs) and hourglass nodal-line semimetals (HNLSMs). They qualitatively differ from 3D Weyl semimetals\cite{volovik-2009,wan-2011, xu-2011,burkov-2011,yang-2011, halasz-2012,zhang-2014, liu-2014PRB,weng-2015, huang-2015,hirayama-2015,lv-2015a,xu-2015a,yang-2015,lv-2015b,xu-2015b,lu-2015, ruan-2016NC,ruan-2016PRL,wang-2016PRA,lian-2016PRB,soluyanov-2015,yan-2015prb,deng-2016NP,liang-2016,ding-2016,jiang-2017NC,young-2012,wang-2012,wang-2013,steinberg-2014,liuzk-2014a,liuzk-2014b,borisenko-2014,neupane-2014,yang-2014} and the nodal-line semimetals\cite{burkov-2011b,carter-2012,phillips-2014,chen-2015,zeng-2015,chiu-2014,mullen-2015,weng-2015b,yu-2015,kim-2015, bian-2016,xie-2015,rhim-2015,chen-2015b,fang-2015b,bian-2016b,chan-2015,bzdusek-2016} studied before in that they have quadruplets of bands with internal partner-switching. The HWSMs host four Weyl points at each screw-invariant momentum line, while the HNLSMs have nodal lines in glide-invariant momentum planes. Note that the quadruplet bands studied in the present work are closely related to filling-enforced band insulators in which nonsymmorphic symmetries lead to tighter filling constraints\cite{parameswaran-2013,watanabe-2015,po-2016,watanabe-2016, parameswaran-2015, pixley-2016}.

{\it Hourglass semimetals in lower dimensions}.---Hourglass-like dispersions in 2D bulk systems protected by nonsymmorphic symmetries were studied in Refs.\cite{young-2015,wieder-2016PRB}. Here, in order to construct explicit 3D models with hourglass-like bulk dispersions, we first consider 2D hourglass semimetals and then properly stack them into 3D systems with hourglass-like dispersions in the 3D bulk bands.

\begin{figure}[t]
\centering
\includegraphics[width=6.0cm]{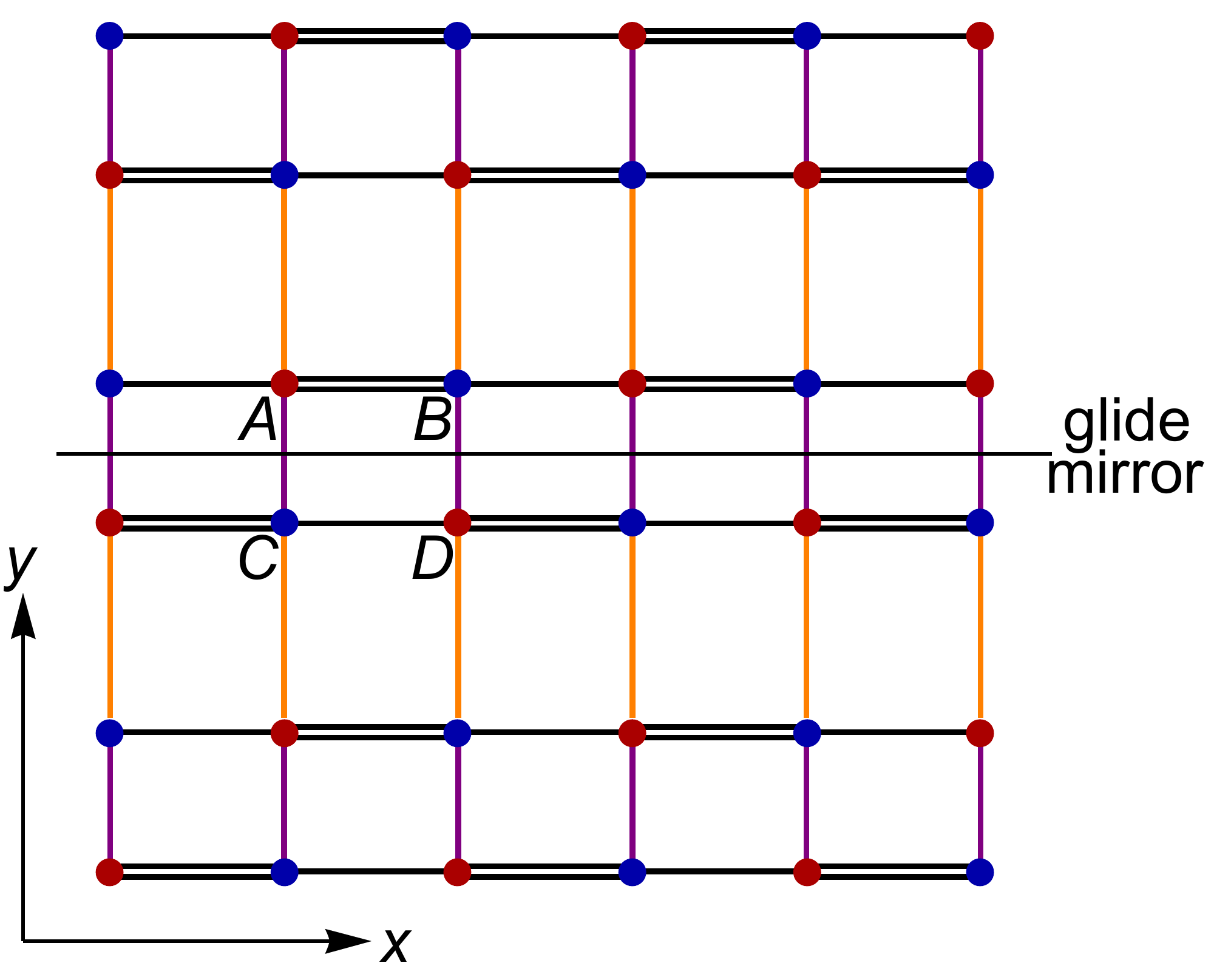}
\caption{The lattice structure of the 2D model. The lattice is obtained by stacking ladders, each of which consists of two nearest neighbor horizontal chains. Within each horizontal chain, the black double lines represent the hopping $t+\delta t$ and the black lines represent the hopping $t-\delta t$. The purple lines indicate the hopping and spin-orbit coupling between the chains within the ladder, and the orange lines indicate the hopping and spin-orbit coupling between neighboring ladders. Each unit cell has four inequivalent sites $A, B, C$ and $D$.}\label{2D}
\end{figure}

The 2D lattice can be constructed by stacking two-leg ladders with a glide reflection symmetry indicated by the long black line in Fig. \ref{2D}. More explicitly, the lattice remains invariant when reflected by the glide mirror and translated by half unit cell along the $x$-axis. Each ladder consists of two neighboring Su-Schrieffer-Heeger (SSH) chains\cite{su-1979} having two sublattices indicated by the red and blue dots. Within each chain, the hoppings between the two sublattices have amplitude $t+\delta t$ (black double lines) and $t-\delta t$ (black single lines), respectively. And the purple and orange lines represent hopping between neighboring chains. The unit cell has four inequivalent sites $A$, $B$, $C$ and $D$, as shown in Fig. \ref{2D}.

We first write down the following Hamiltonian describing a ladder while neglecting its coupling with other ladders:
\begin{eqnarray}\label{H1D}
&&H_{l} = \sum_{n=1}^{N}\bigg\{\sum_{s=\uparrow,\downarrow}\left[(t+\delta t)c_{As n}^\dagger  c_{Bs n}+(t-\delta t)c_{Bs n}^\dagger c_{As n+1}\right.\nonumber \\
&&+ (t-\delta t)c_{Cs n}^\dagger c_{Ds n}+(t+\delta t)c_{Ds n}^\dagger c_{Cs n+1}\nonumber\\
&&+ \left.t'(c_{As n}^\dagger c_{Cs n}+c_{Bs n}^\dagger c_{Ds n}) \right]\nonumber\\
&&- \lambda(c_{A\uparrow n}^\dagger c_{C\downarrow n}\!-\!c_{A\downarrow n}^\dagger c_{C\uparrow n}\!-\!c_{B\uparrow n}^\dagger c_{D\downarrow n}\!+\!c_{B\downarrow n}^\dagger c_{D\uparrow n})\!+\!h.c.\bigg\},~~~
\end{eqnarray}
where $c_{is n}$ annihilates an electron with spin $s$ at the $i$-site ($i=A, B, C$ or $D$) of the $n$-th unit cell, and periodic boundary condition $c_{is N+1}=c_{is 1}$ is assumed. $t'$ is the interchain hopping within the ladder and $\lambda$ is the strength of spin-orbit coupling. By performing Fourier transformations, we have $H_{l} = \sum_{k_x}\Psi^\dagger(k_x) h_{l}(k_x)\Psi(k_x) $, where $\Psi=(c_{As }, c_{Bs}, c_{Cs }, c_{Ds })$ and $h_{l}(k_x)$ reads
\begin{eqnarray}\label{h1d}
h_{l}(k_x) &=& t(1+\cos k_x)\s_x+t\sin k_x \s_y+\delta t(1-\cos k_x)\s_x\tau_z\nonumber\\
  &&-\delta t\sin k_x \s_y\tau_z+t' \tau_x+\lambda s_y\s_z\tau_y,
\end{eqnarray}
where $s_\alpha$, $\s_\alpha$ and $\tau_\alpha$ ($\alpha=x,y,z$) are Pauli matrices acting on spin, sublattice and chain subspace, respectively.

The ladders are coupled through inter-ladder hopping, as indicated by the orange lines in Fig. \ref{2D}. The resulting 2D Hamiltonian reads
\begin{eqnarray}
h_{2D}(k_x,k_y) &=&h_{l}(k_x)+t_y(\cos k_y\tau_x-\sin k_y\tau_y)\nonumber\\
 &&+\lambda_y(\cos k_ys_y\s_z\tau_y+\sin k_y s_y\s_z\tau_x),\label{h2D}
\end{eqnarray}
where $t_y$ and $\lambda_y$ are the amplitudes of the inter-ladder hopping and spin-orbit coupling, respectively. The system respects time-reversal symmetry, with time reversal operator $T=is_yK$, with $K$ being the complex conjugate operator. It also respects the following glide symmetry: $G_y(k_x)h_{2D}(k_x,k_y)G_y(k_x)^{-1} = h_{2D}(k_x,-k_y)$, where the glide reflection operator is
\begin{eqnarray}\label{glide}
G_y(k_x) &=& is_y\otimes\left(\begin{array}{cc}
                             0 & e^{-ik_x} \\
                             1 & 0
                           \end{array}\right)_\s\otimes\tau_x.
\end{eqnarray}
Here, $is_y$ represents the spin rotation under the reflection, $\tau_x$ interchanges $A$ and $C$ as well as $B$ and $D$ sites and the matrix with $\sigma$ indices represents the effect of translation by half unit cell along the $x$-direction.

\begin{figure}[t]
    \subfigure[]{\includegraphics[height=3.3cm]{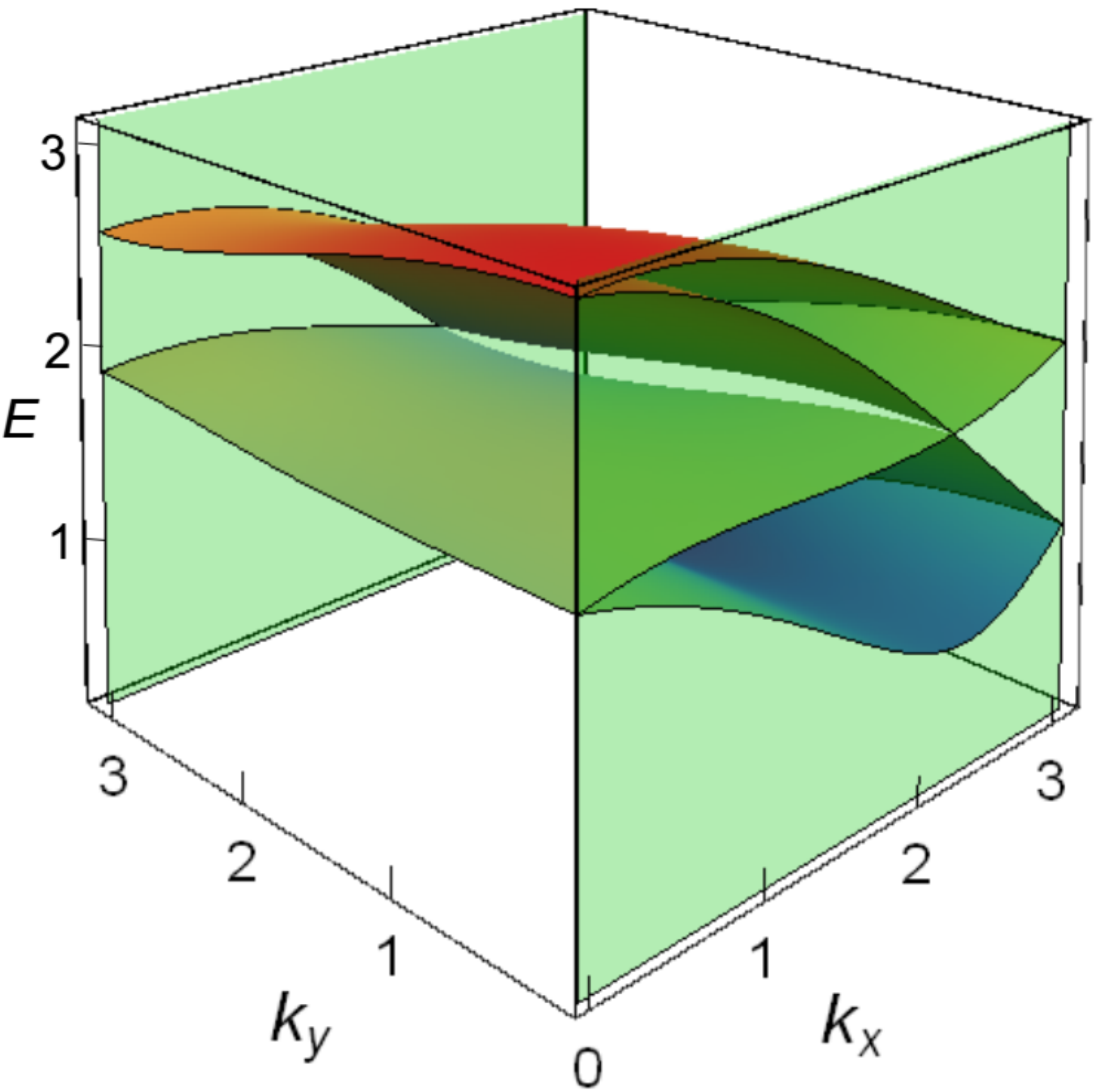}\label{2Dband1}}~~~~~~
    \subfigure[]{\includegraphics[height=3.cm]{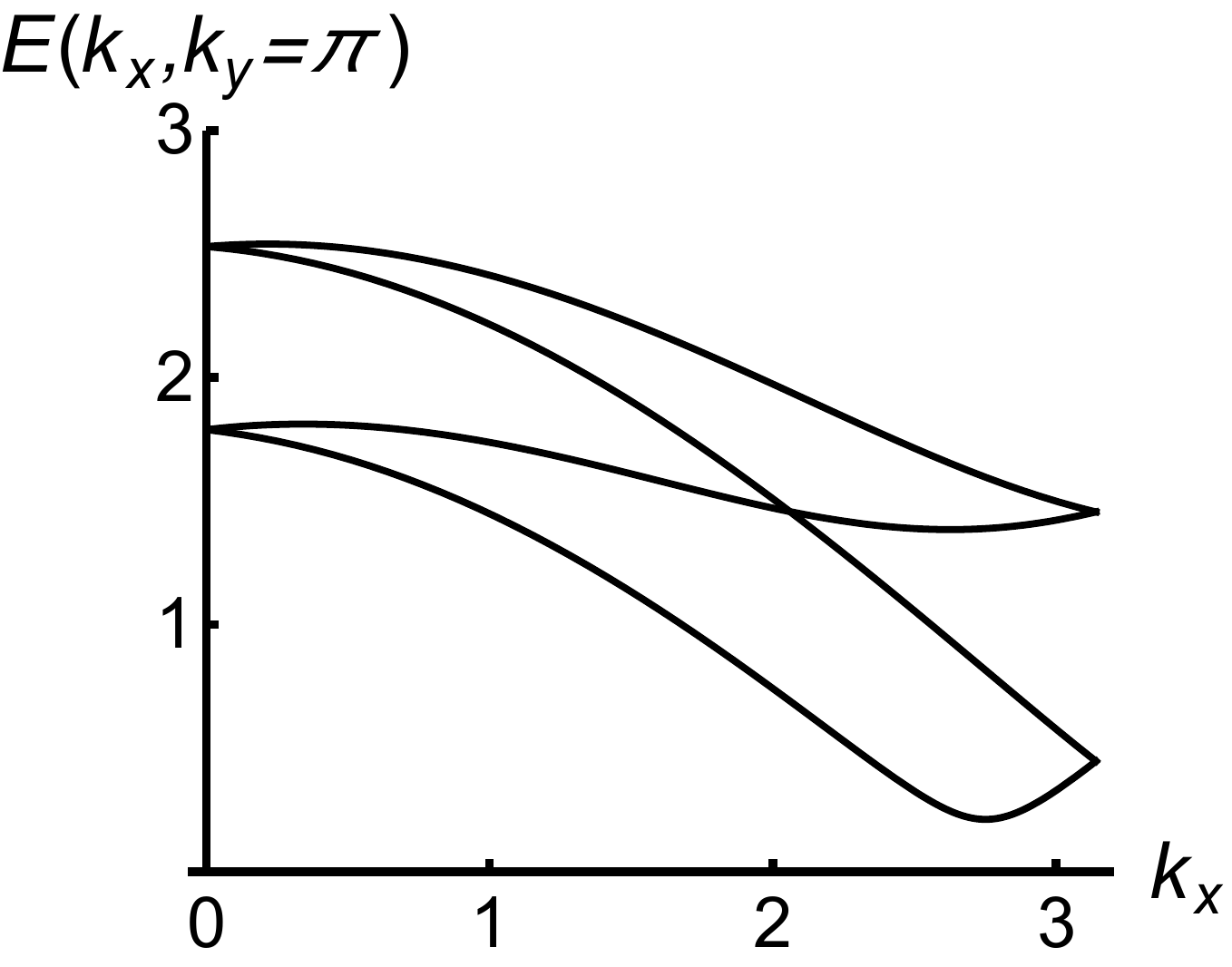}\label{2Dband1pi}}
   \caption{(a) The band structure of the 2D lattice. (b) The dispersion along $k_y=\pi$. Only the quadruplet with positive energy is shown. The parameters are $\delta t=0.3t, t'=0.6t, t_y=0.2t, \lambda=t,\lambda_y=0.2t$.}
\end{figure}

At each time-reversal invariant momentum (TRIM) $(\bar k_x,\bar k_y)$ where $\bar k_x, \bar k_y=0,\pi$, the spectrum are Kramers-degenerate. There are two glide-invariant lines, $k_y=\bar k_y$ where we can simultaneously diagonalize $h_{2D}(k_x,\bar k_y)$ and $G_y(k_x)$, and then label each band by the eigenvalue of $G_y(k_x)$, i.e. the glide parity $g_\pm=\pm ie^{-i\frac{k_x}{2}}$. Therefore, at $(0,\bar k_y)$, two bands with glide parity $g_+$ and $g_-$ are degenerate, while at $(\pi,\bar k_y)$, two bands with the same glide parity are degenerate. This results in band crossing at $(k_{x}^*,\bar k_y)$ between two TRIMs, which is protected by time-reversal and glide symmetries.

The band structure is shown in Fig. \ref{2Dband1}, and, in particular, the hourglass-like structure along $k_y=\pi$ is shown in Fig. \ref{2Dband1pi}. (The eight bands split into two quadruplets with a gap between them, each of which displays hourglass-like dispersion, and we only show the upper quadruplet.) The double degeneracy in dispersions along $k_x=\pi$ is due to a combined symmetry $TG_y(k_x)$, such that at $k_x=\pi$, $(TG_y(\pi))^2=-1$. But, the double degeneracy at a generic momentum along the $k_x=0$ line is not protected by symmetries; actually other symmetry-preserving spin-orbit coupling, such as the term $\lambda_y'\sin k_y s_x\s_z\tau_z$, can split this degeneracy. For convenience, we call the crossings at $k_x=0, k^\ast_{x}$, and $\pi$ the ``inner edge", the ``neck" and the ``outer edge" of the hourglass, respectively.

In fact, hourglass-like dispersion can also be realized in 1D systems which respect time-reversal and glide or screw rotation symmetries. The ladder Hamiltonian given by Eq.(\ref{h1d}) is an example. We further construct a minimal model for 1D hourglass fermions in the Supplemental Material\cite{SM}. Below, we shall focus on hourglass fermions in bulk band structures of 3D systems.

{\it Hourglass semimetals in three dimensions}.---Here, we introduce hourglass semimetals in 3D, which have hourglass-like dispersion along at least one line in momentum space. Depending on the dimension of the manifold of degeneracies, we divide the hourglass semimetals into hourglass Weyl semimetals (HWSMs) and hourglass nodal-line semimetals (HNLSMs).

\begin{figure}[t]
  \centering
  \subfigure[]{\includegraphics[height=2.6cm]{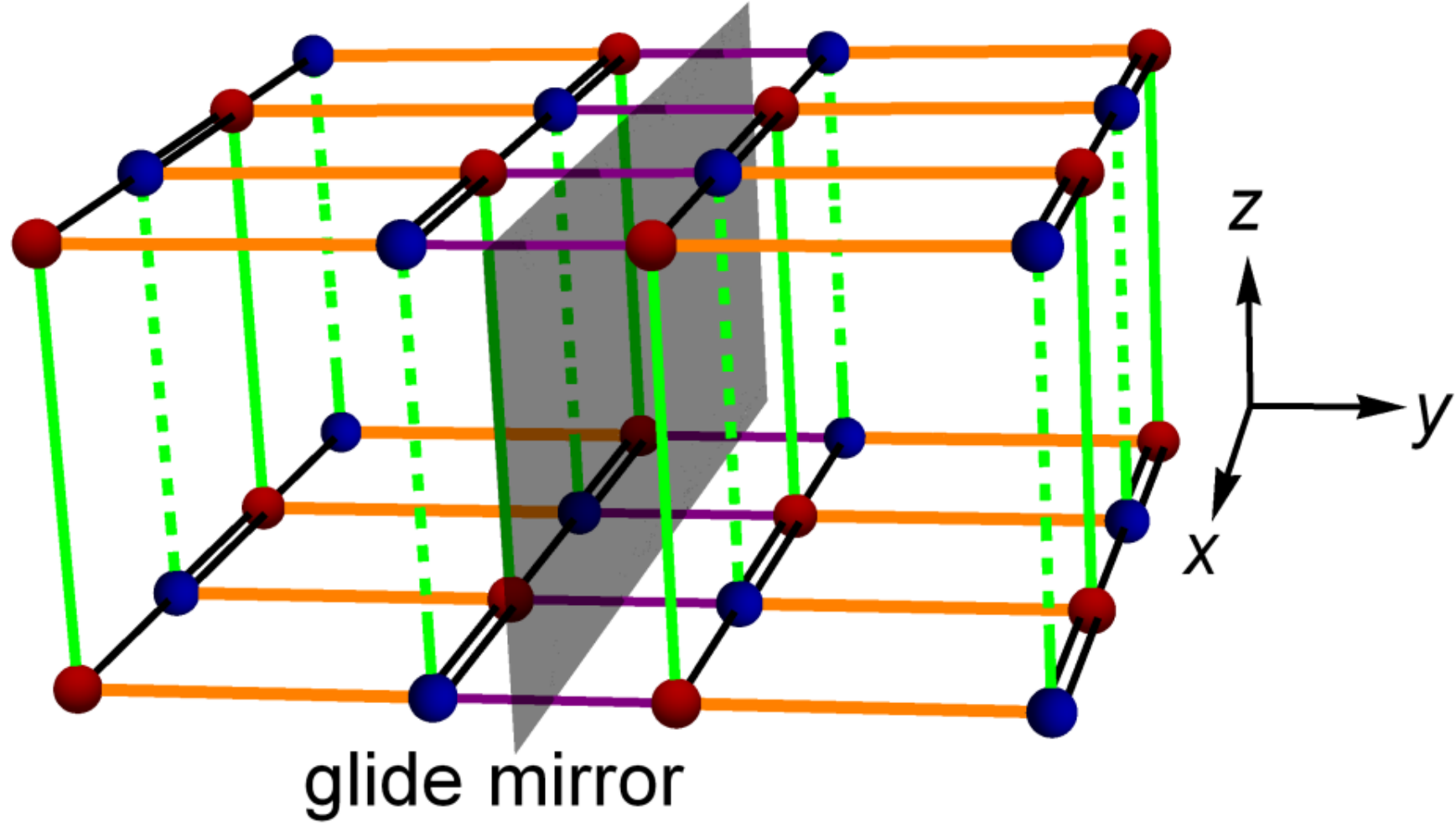}\label{NLlattice}}~
  \subfigure[]{\includegraphics[height=3.0cm]{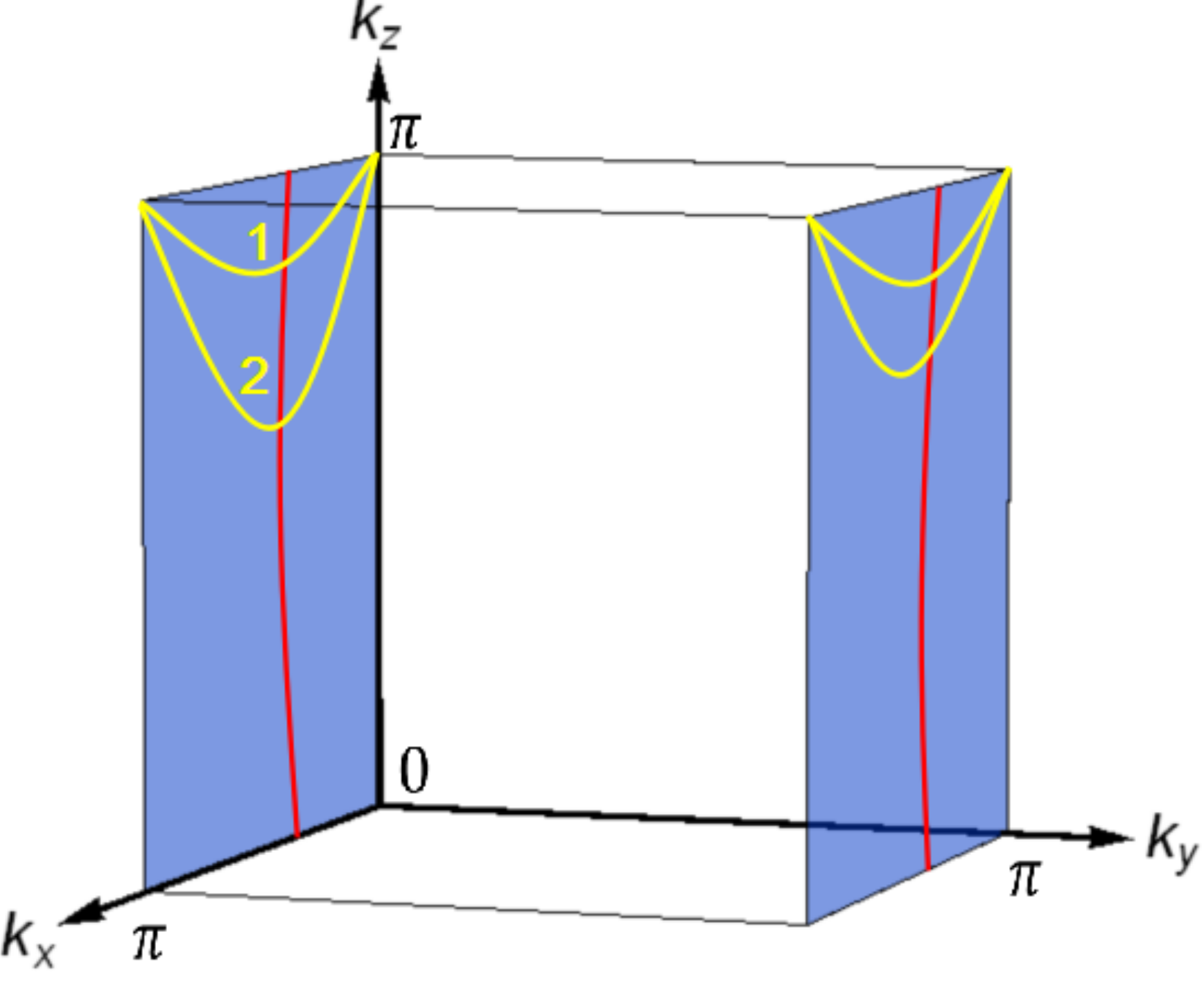}\label{BZNL}}
  \subfigure[]{\includegraphics[height=3.cm]{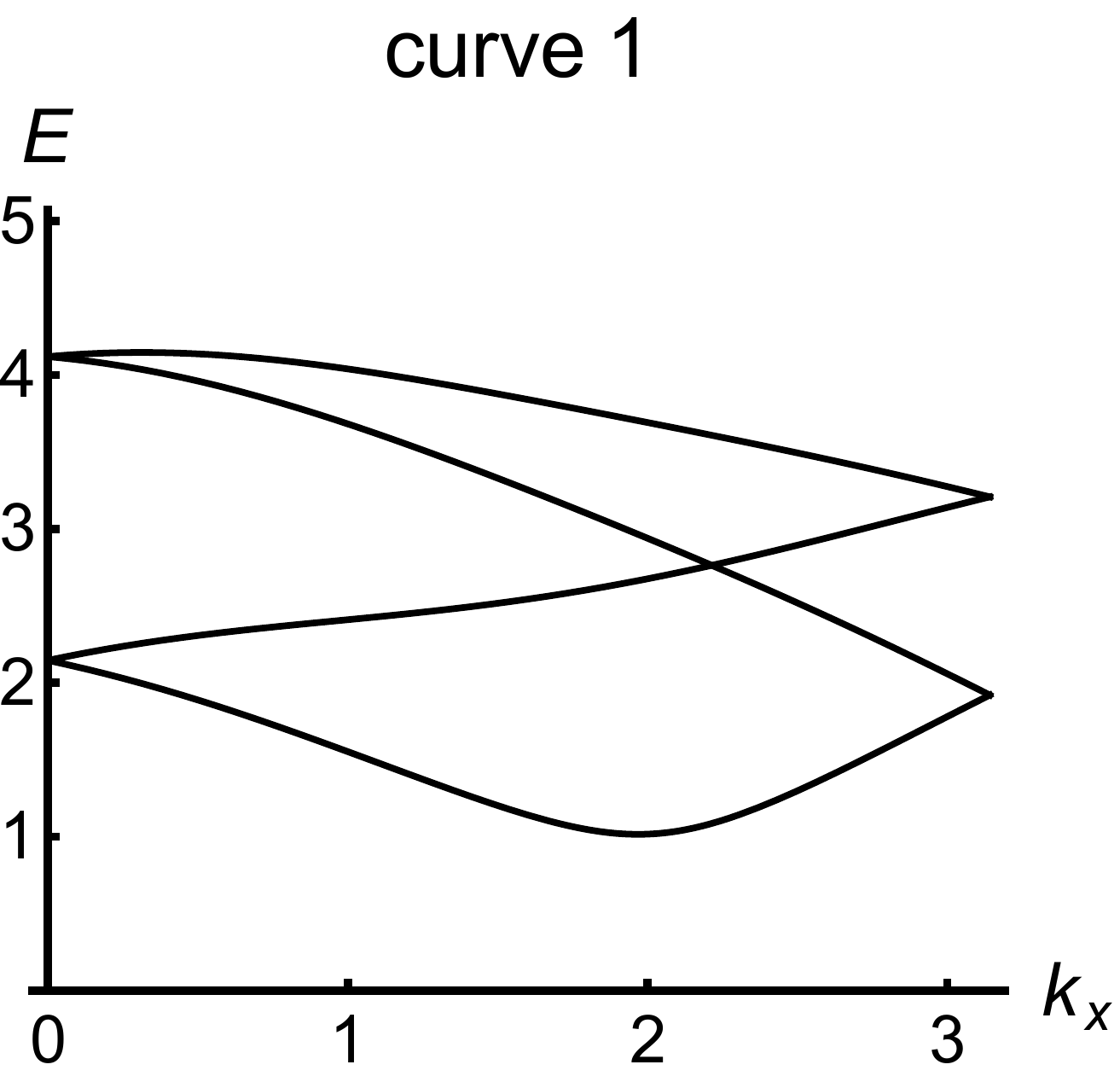}\label{3DNL1}}~~~~
  \subfigure[]{\includegraphics[height=3.cm]{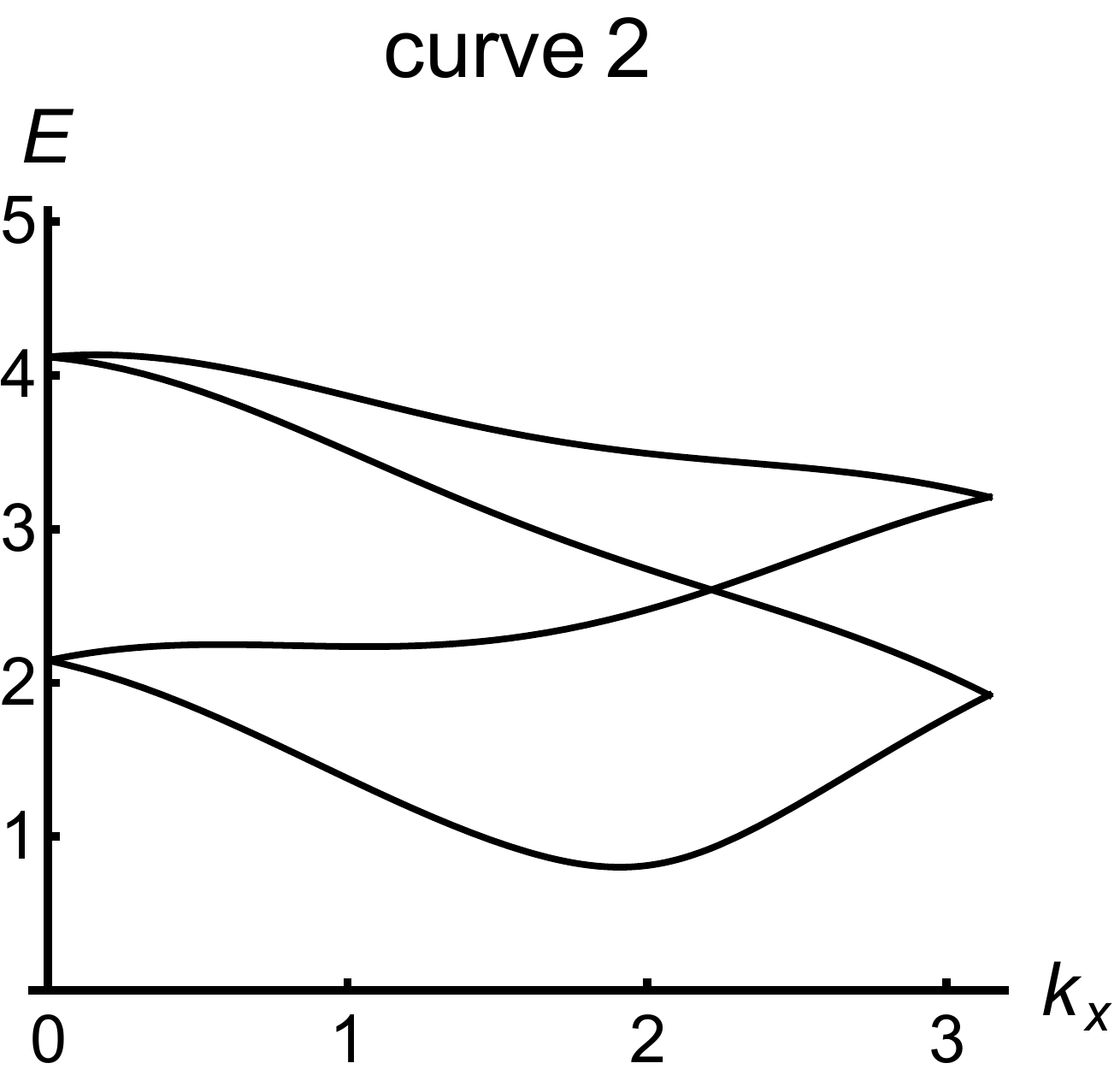}\label{3DNL2}}
  \caption{(a) Layer construction of the hourglass nodal-line semimetal. The hopping along the $z$-axis is $t_{z1}$ for $A$ and $D$ sublattices (green lines) and $t_{z2}$ for $B$ and $C$ sublattices (green dashed lines). (b) The Brillouin zone of hourglass nodal-line semimetals, with the red lines representing the nodal lines. The blue planes are glide-invariant planes. Along any curve in the plane connecting $(0,\bar k_y,\bar k_z)$ and $(\pi,\bar k_y,\bar k_z)$ (such as the yellow curves), the dispersion is hourglass-like. (c), (d) The hourglass-like dispersion along two specific curves shown in (b). The parameters are $\delta t=0.5t, t'=0.5t, t_y=0.7t, \lambda=0.8t,\lambda_y=0.4t,\lambda_y'=0.3t,t_{z1}=0.6t,t_{z2}=0.8t,\lambda_z=0.3t$.}\label{3D}
\end{figure}

We construct a 3D model featuring HNLSMs by stacking the 2D layers described in Eq.(\ref{h2D}) along the $z$-axis, as shown in Fig. \ref{NLlattice}. The resulting 3D Hamiltonian with spin-orbit couplings is given by
\begin{eqnarray}
&&h_{3D}(k_x,k_y,k_z)=h_{2D}(k_x,k_y)+\frac{1}{2}(t_{z1}+t_{z2})\cos k_z\nn\\
&&+\frac{1}{2}(t_{z1}-t_{z2})\cos k_z\s_z\tau_z+\lambda_{y}'\sin k_ys_x\s_z\tau_z\nonumber\\
&&+\lambda_{z}\sin k_z s_y\s_z\tau_z,
\end{eqnarray}
which respects both time-reversal and glide symmetries, where the glide reflection operator $G_y(k_x)$ is given by Eq. (\ref{glide}).

In the glide-invariant plane $k_y=\bar k_y$, each band can be labeled by the glide parity, which is well defined. Again, the Kramers partners at $(0,\bar k_y,\bar k_z)$ have different glide parity and those at $(\pi,\bar k_y,\bar k_z)$ have the same glide parity, where $\bar k_z=0,\pi$. Therefore, along any curve in the $k_y=\bar k_y$ planes connecting $(0,\bar k_y,\bar k_z)$ and $(\pi,\bar k_y,\bar k_z)$, the dispersion is hourglass-like. The dispersions along two of those curves are shown in Figs. \ref{3DNL1} and \ref{3DNL2}. The necks are joined to a nodal line, as shown in Fig. \ref{BZNL}. Note that similar nonsymmorphic nodal-line semimetals were studied in Ref.\cite{bzdusek-2016}.

\begin{figure}[t]
  \centering
  \subfigure[]{\includegraphics[width=4cm]{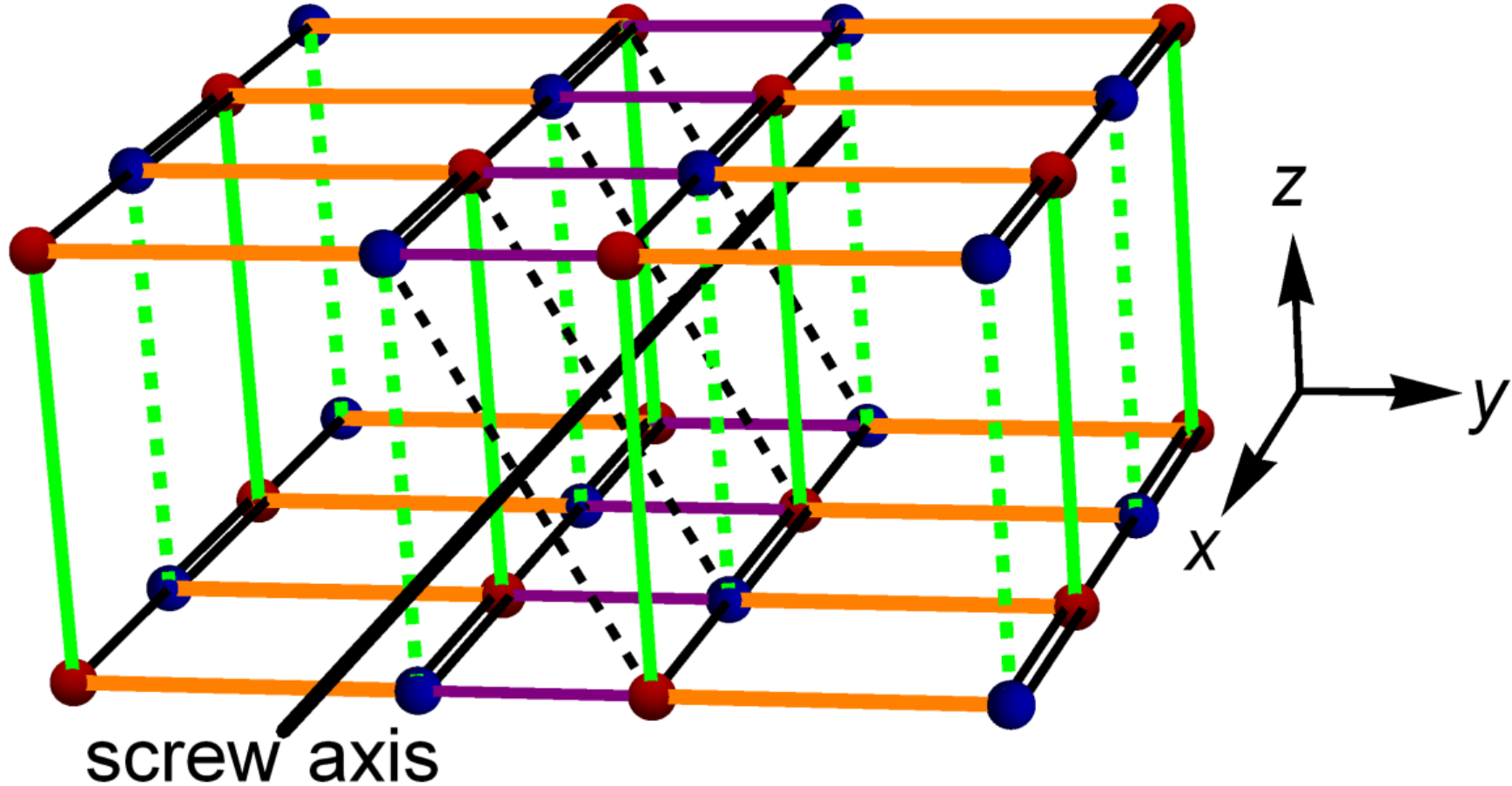}\label{3DWP}}~~
  \subfigure[]{\includegraphics[width=3.5cm]{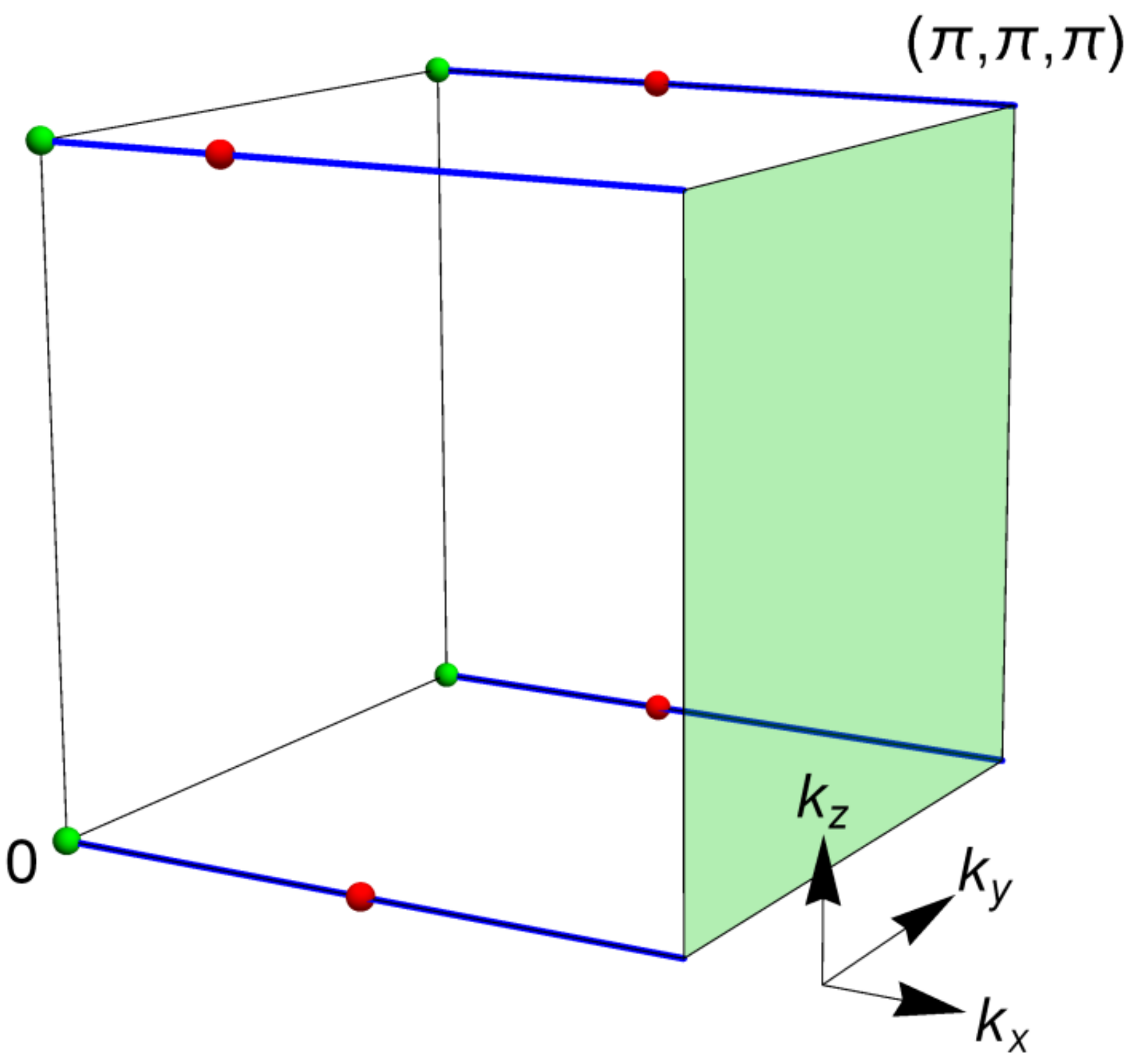}\label{BZWP}}
  \subfigure[]{\includegraphics[width=4.cm]{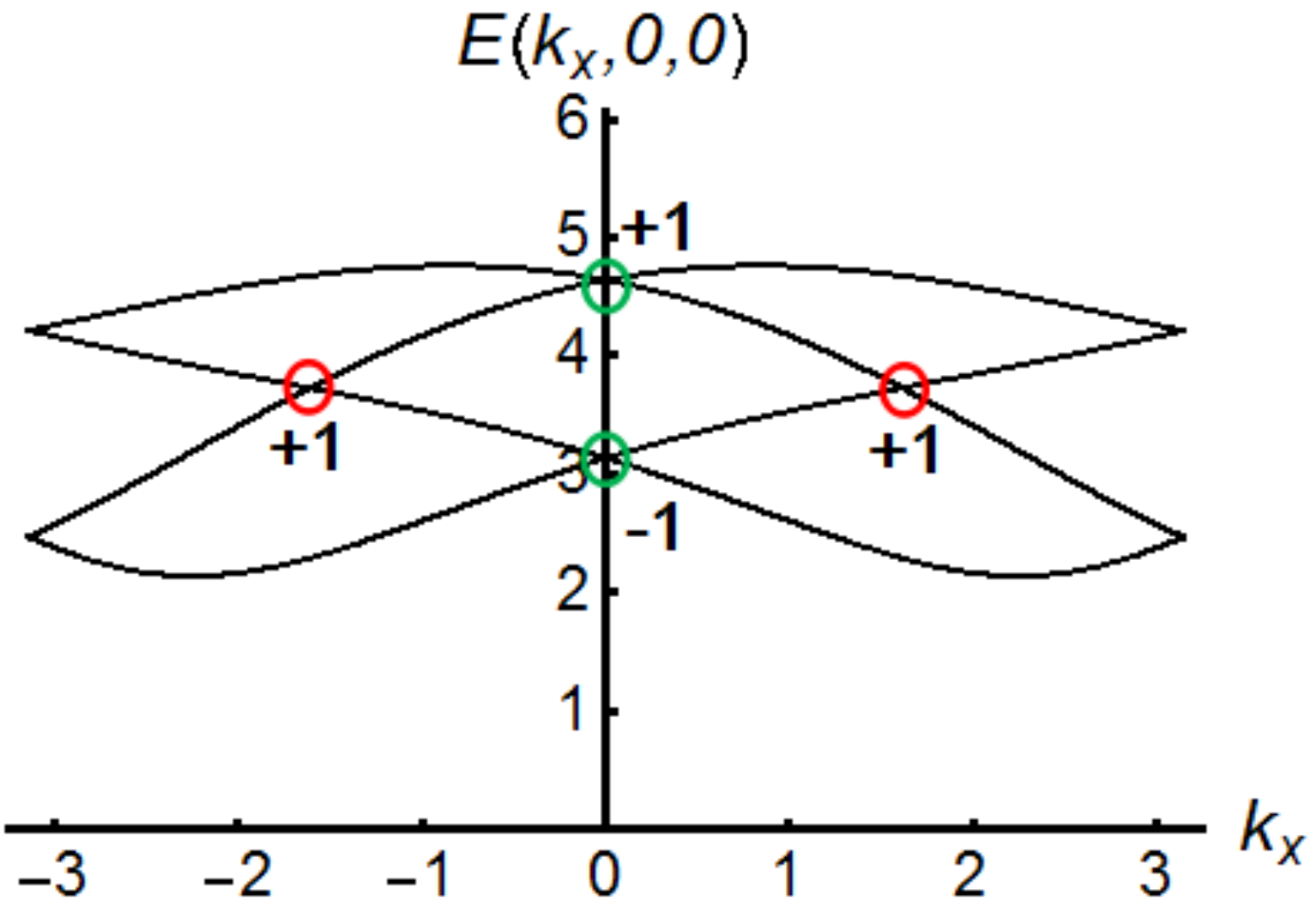}\label{3DWeyl}}~~
  \subfigure[]{\includegraphics[width=4.cm]{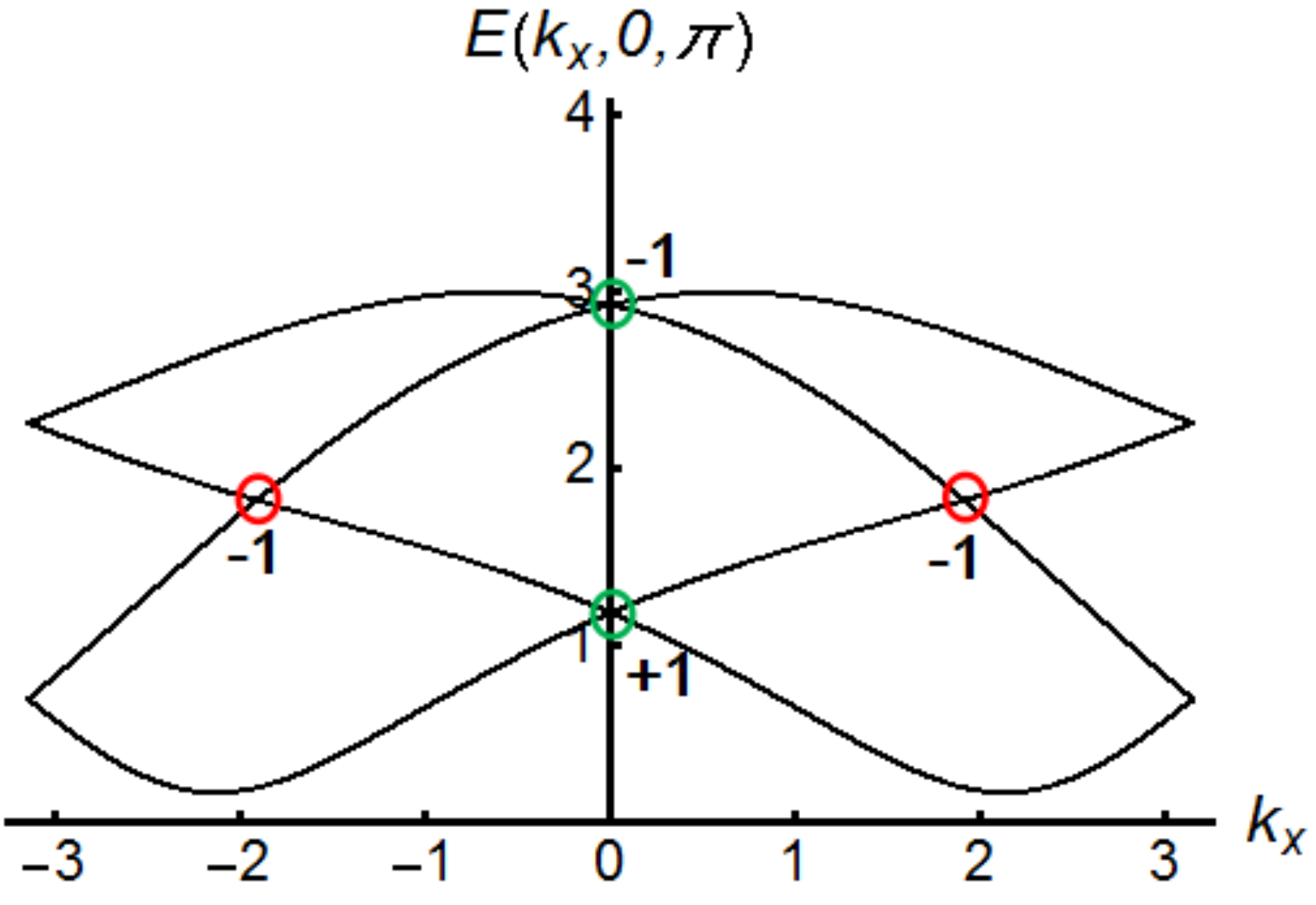}\label{3DWeyl1}}
  \caption{(a) Layer construction of the hourglass Weyl semimetal. Spin-orbit coupling between $A$($B$) sites in one layer and $C$($D$) sites in the layer above is indicated by the black dashed lines. (b) The Brillouin zone of hourglass Weyl semimetals, with the red points indicating Weyl points at the neck of the hourglasses, the green points indicating Weyl points at the inner edge, the blue lines indicating screw-invariant lines, and the green plane indicating double degeneracy in the whole plane. (c) The dispersion along the $k_x$-axis. (d) The dispersion along $(k_x,0,\pi)$. In (c) and (d), the Weyl points at the neck and the inner edge of the hourglasses are marked by red and green circles, respectively, with their monopole charges labeled. $\lambda=1.6t,\lambda_{z1}=\lambda_{z2}=\lambda_{z3}=0.3t$, and the other parameters are the same as in Fig.\ref{3D}.}\label{3D2}
\end{figure}

Although the glide symmetry can only protect the HNLSMs, we now show that an effective screw symmetry, whose transformation is a combination of the screw rotation with spin rotation, can protect a HWSM in which the hourglass-like dispersion only appears along the high symmetry momentum lines $k_y=\bar k_y, k_z=\bar k_z$. For example, we consider the following Hamiltonian in 3D,
\bea
&&h'_{3D}(k_x,k_y,k_z)= h_{2D}(k_x,k_y)+\frac{1}{2}(t_{z1}+t_{z2})\cos k_z\nonumber\\
 &&+\frac{1}{2}(t_{z1}-t_{z2})\cos k_z\s_z\tau_z+\lambda_{z1}\sin k_zs_z\s_z\tau_z+\lambda_{z2}\sin k_zs_y\tau_z\nn\\
 &&+\lambda_{z3}(\cos k_zs_y\s_z\tau_y+\sin k_zs_y\s_z\tau_x)+\lambda_{y}'\sin k_ys_x\s_z\tau_z.
\eea
It is clear that the Hamiltonian $h'_{3D}$ respects a combined symmetry $(is_z S_x)h'_{3D}(k_x,k_y,k_z)(is_z S_x)^{-1} = h'_{3D}(k_x,-k_y,-k_z)$, where
\begin{eqnarray}\label{screw}
  S_x &=& is_x\otimes\left(\begin{array}{cc}
                             0 & e^{-ik_x} \\
                             1 & 0
                           \end{array}\right)_\s\otimes\tau_x,
\end{eqnarray}
as well as time-reversal symmetry. The screw rotation $S_x$ is illustrated from Fig. \ref{3DWP}: the rotation around the screw axis indicated by the black line followed by a half unit cell translation along the $x$-axis leaves the lattice invariant.

At the screw-invariant lines $k_y=\bar k_y,k_z=\bar k_z$, each band can be labeled by the eigenvalue of $S_x$, which is $\pm ie^{-i\frac{k_x}{2}}$. 
Analogous to the same argument in the case of a glide symmetry, the dispersion along $k_x$ on these high-symmetry lines $k_y=\bar k_y,k_z=\bar k_z$ is hourglass-like. The necks of the hourglasses are Weyl points. Moreover, there are two Weyl points at the inner edge ($k_x=0$) of the hourglass, while the outer edge is in the $k_x=\pi$ plane where double degeneracy occurs in the whole plane\cite{liang-2016PRB} due to the antiunitary symmetry $Tis_zS_x(\pi)$ with $(Tis_zS_x(\pi))^2=-1$. Therefore, there are four Weyl points on each screw-invariant line, and 16 Weyl points in total. The degeneracies are schematically shown in Fig. \ref{BZWP}, and the dispersions along the $k_x$-axis and the line $(k_x,0,\pi)$ are shown in Figs. \ref{3DWeyl} and  \ref{3DWeyl1}.

\begin{figure}[t]
  \centering
  \subfigure[]{\includegraphics[width=3.8cm]{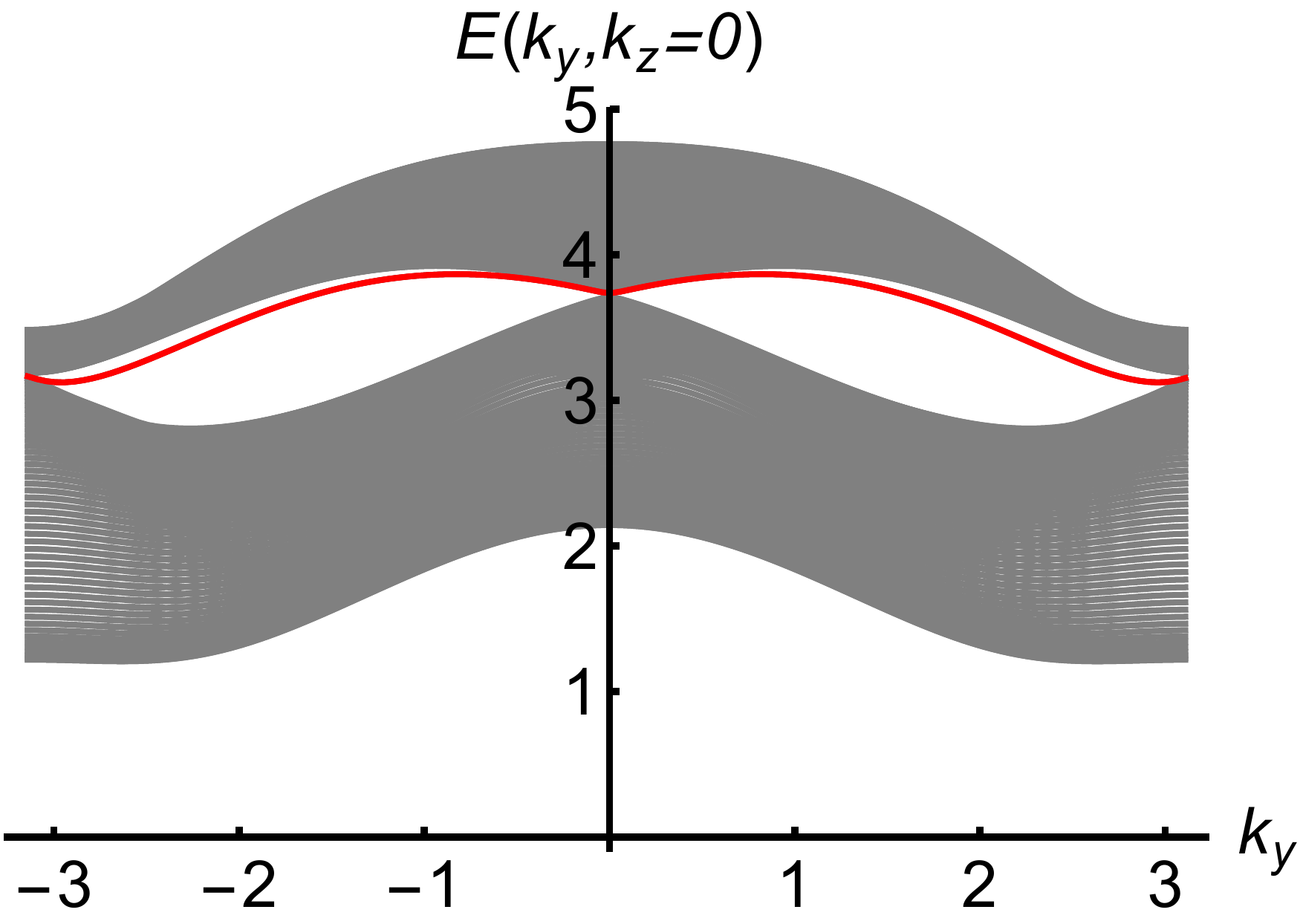}\label{surface}}~~
  \subfigure[]{\includegraphics[width=3.8cm]{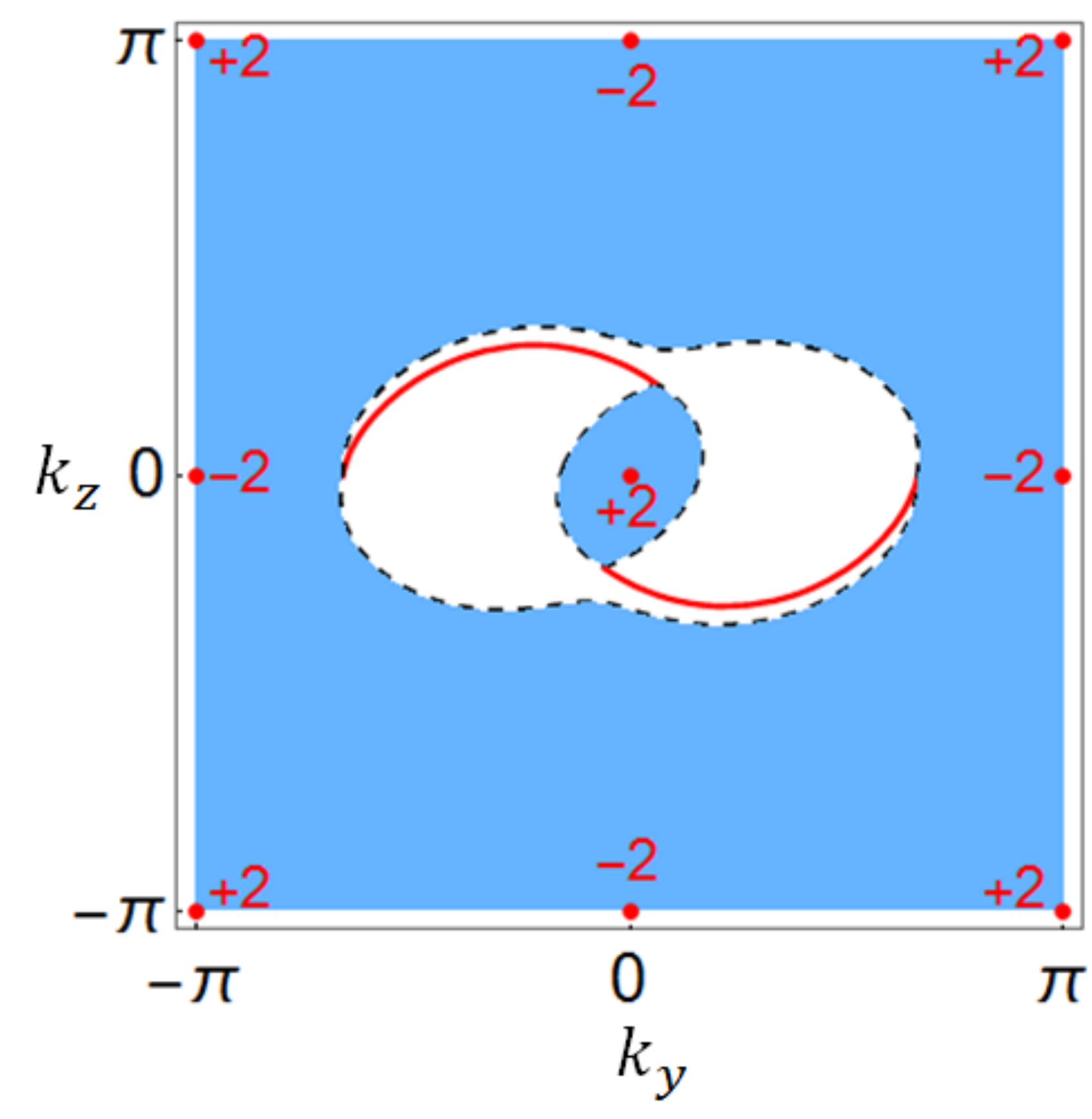}\label{FermiArc}}
  \caption{(a) The (100) surface state along the $k_y$-axis in the surface Brillouin zone. The bulk bands projected on the surface are in gray, while the surface bands are in red. (b) The surface Fermi pockets (black dashed contours) enclosing the Weyl points with opposite charge projected on the surface are connected by Fermi arcs (red curves). The parameters are the same as in Fig. \ref{3DWeyl}, and the Fermi energy is assumed to be $3.5t$.}
\end{figure}

We would like to emphasize that Weyl points on the screw-invariant lines are topologically stable. Namely, even if perturbations that break the screw rotation symmetry are introduced, the hourglass structures still exist as long as the Weyl points do not meet each other, due to the topological nature of monopole charges associated with each Weyl point. The edges of the hourglasses do not shift due to time-reversal symmetry, but their neck can shift away from the screw-invariant lines when weak screw-symmetry-breaking perturbations are present. Furthermore, the outer edge of the hourglasses can become Weyl points, since the double degeneracy in the plane $k_x=\pi$ can be lifted by the perturbations except at the TRIMs.

{\it Surface states}.---On the surface of the HWSMs, surface Fermi arcs connecting the projected Weyl points are expected. We show the surface states on (100) surface in Fig. \ref{surface} along the $k_y$-axis. The surface arcs, as indicated by the red curves, are separated from the projected bulk bands in gray, except at the Weyl points where they connect. Since the Weyl points not related by symmetries are generically not at the same energy, two Fermi pockets, each enclosing one of the projected Weyl points, occur when the Fermi energy is between the energy of two Weyl points. The two pockets are connected by Fermi arcs, as shown in Fig. \ref{FermiArc}, in which $+2$ and $-2$ indicate the total monopole charge of the Weyl points that are projected to the surface Brillouin zone. The Fermi arc could be experimentally observed by angle-resolved photoemission spectroscopy (ARPES) experiments.

{\it Conclusion}.---In conclusion, we have proposed models in 3D that host fermions with hourglass-like dispersion along high symmetry lines or curves in high symmetry planes in systems respecting nonsymmorphic symmetries. The hourglass-like dispersion is protected by time-reversal symmetry and glide reflection or screw rotation symmetry. In 3D, depending on the type of nonsymmorphic symmetry, either HWSMs or HNLSMs could be realized. 3D hourglass semimetals with nonsymmorphic symmetries, if discovered in solid state materials, could provide a new arena to explore novel physics in Weyl semimetals and nodal-line semimetals.

{\it Acknowledgement}. L.W. thanks C. Fang and Z. Yan for useful discussions. This work was in part supported by the NSFC under Grant No. 11474175 (LW, SKJ, HY) and by the Ministry of
Science and Technology of China under Grant No. 2016YFA0301001 (HY).

\begin{widetext}
\section{supplemental material}
\renewcommand{\theequation}{S\arabic{equation}}
\setcounter{equation}{0}
\renewcommand{\thefigure}{S\arabic{figure}}
\setcounter{figure}{0}
\renewcommand{\thetable}{S\arabic{table}}
\setcounter{table}{0}

\subsection{The four-band model for 1D hourglass fermion}
In this supplemental materials, we construct a minimal four-band model featuring 1D hourglass fermions. Consider a zigzag chain as shown in Fig. S\ref{zigzag}. It respects glide reflection symmetry indicated by the dashed line. Glide reflection transformation, i.e., reflection followed by translating half lattice constant, in sublattice space reads
\bea
\left(\ba{cccc} 1 & 0 \\ 0 & e^{ik_x} \ea \right) \left( \ba{cccc} 0 & 1 \\ 1 & 0 \ea \right)= \left( \ba{cccc} 0 & 1 \\ e^{ik_x} & 0 \ea \right)= e^{i \frac{k_x}{2}}(\cos \frac{k_x}{2} \sigma_x+ \sin \frac{k_x}{2} \sigma_y),
\eea
Let $\sigma_\parallel(k_x)=\cos \frac{k_x}{2} \sigma_x+ \sin \frac{k_x}{2} \sigma_y$, the symmetry transformation including spin space reads
\bea
	G_y(k_x)= i e^{i \frac{k_x}{2}}s_y \otimes \sigma_\parallel(k_x),
\eea
One can introduce anther matrix that are perpendicular to $\sigma_\parallel$, i.e., $\sigma_\perp(k_x)=\cos \frac{k_x}{2} \sigma_y- \sin \frac{k_x}{2} \sigma_x$. Note that $[T, \sigma_\parallel(k_x)]=0,\{T, \sigma_\perp(k_x)\}=0$. The most general time-reversal invariant Hamiltonian describing nearest neighbor hopping or spin-orbit coupling is given by,
\bea
	H= -t \cos \frac{k_x}{2} \sigma_\parallel(k_x)+\lambda_1 \sin \frac{k_x}{2} s_y \sigma_\parallel(k_x)+ \lambda_2 \cos \frac{k_x}{2} s_x\sigma_\perp(k_x). \label{4band_ham}
\eea
The dispersion is shown in Fig. S\ref{4band} with plotting parameter $t=\lambda_1=1,\lambda_2=0$, i.e., the first two terms are enough for hourglass-like dispersion. The hourglass-like dispersion is protected by time-reversal symmetry and glide reflection symmetry.

\begin{figure}[h]
	\subfigure[]{\label{zigzag}
		\includegraphics[width=4.5cm]{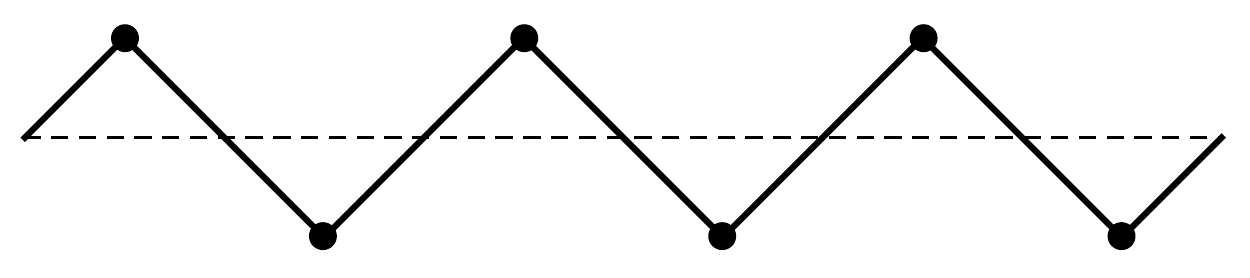}}~~~~~~~~~~
	\subfigure[]{\label{4band}
		\includegraphics[width=4.5cm]{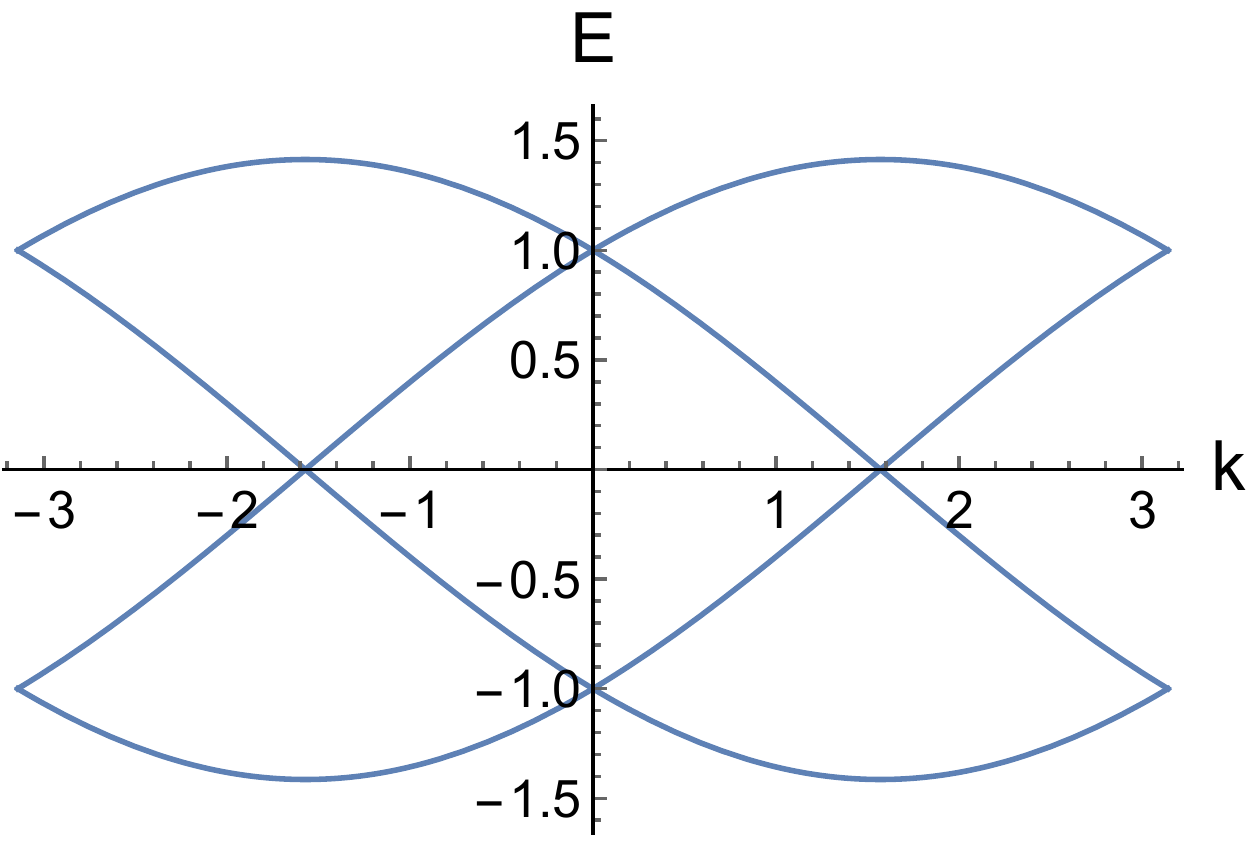}}
\caption{(a) The zigzag chain where the dashed line indicates the glide mirror plane. (b) The dispersion of Hamiltonian Eq. (\ref{4band_ham}). The plotting parameters are $t=\lambda_1=1, \lambda_2=0$ for simplicity.}
\end{figure}

\end{widetext}

\end{document}